\documentclass[11pt,a4paper]{article}
\usepackage[margin=2.5cm]{geometry}
\usepackage{amsmath,amssymb,graphicx}
\usepackage[dvipsnames]{xcolor}
\usepackage[hidelinks]{hyperref}
\DeclareUnicodeCharacter{2062}{}

\makeatletter
\newcommand{\articletype}[1]{}
\def\@affils{}
\newcommand{\affil}[1]{\g@addto@macro\@affils{#1\par}}
\def\@email{}\newcommand{\email}[1]{\gdef\@email{#1}}
\def\@keywords{}\newcommand{\keywords}[1]{\gdef\@keywords{#1}}
\newcommand{\makeheader}{%
  \begin{center}
    {\LARGE\bfseries \@title\par}\vspace{1em}
    {\large \@author\par}\vspace{1em}
    {\small \@affils}
    \ifx\@email\empty\else{\small E-mail: \@email\par}\fi
  \end{center}
  \ifx\@keywords\empty\else\noindent\textbf{Keywords:} \@keywords\par\medskip\fi}
\newcommand{\ack}[1]{\section*{Acknowledgments}#1}
\makeatother
\begin{document}

\articletype{Paper} %	 e.g. Paper, Letter, Topical Review...

\title{Local electronic structure of the Co$^{2+}$ ions in the disordered  perovskite phase of Ba$_3$CoNb$_2$O$_9$: a temperature-dependent X-ray absorption and magnetic circular dichroism study}

\author{M.~M.~Ferreira-Carvalho$^{1,2,*}$, S.~H.~Chen$^1$, S.~R\"o{\ss}ler$^1$, Fanjun Xu$^{3,4}$, A.~T.~M.~N.~Islam$^{3}$, B.~Lake$^{3,4}$, C.~F.~Chang$^1$, C.~Y.~Kuo$^{5,6}$, S.~M.~Valvidares$^7$, P.~Gargiani$^7$, Z.~Hu$^1$, A.~C.~Komarek$^1$, M.~W.~Haverkort$^8$ and L.~H.~Tjeng$^1$}

\affil{$^1$Max Planck Institute for Chemical Physics of Solids, N{\"o}thnitzer Str. 40, 01187 Dresden, Germany}

\affil{$^2$Institute of Physics II, University of Cologne, Z{\"u}lpicher Stra{\ss}e 77, 50937 Cologne, Germany}

\affil{$^3$Helmholtz-Zentrum Berlin f{\"u}r Materialien und Energie GmbH,Hahn-Meitner-Platz 1, 14109 Berlin, Germany}

\affil{$^4$Institut f{\"u}r Festk{\"o}rperphysik, Technische Universit{\"a}t Berlin, Hardenbergstra{\ss}e 36, D-10623 Berlin, Germany}

\affil{$^5$Department of Electrophysics, National Yang Ming Chiao Tung University, No. 1001, Daxue Rd, 30010 Hsinchu, Taiwan}

\affil{$^6$National Synchrotron Radiation Research Center, 101 Hsin-Ann Road, 30076 Hsinchu, Taiwan}

\affil{$^7$ALBA Synchrotron Light Source, E-08290 Cerdanyola del Vall\`{e}s, Barcelona, Spain}

\affil{$^8$Institute for Theoretical Physics, Heidelberg University, Philosophenweg 19, 69120 Heidelberg, Germany}

\affil{$^*$Author to whom any correspondence should be addressed.}
\email{miguel.carvalho@cpfs.mpg.de}

\makeheader
\begin{abstract}
We performed temperature-dependent X-ray absorption spectroscopy (XAS) and X-ray magnetic circular dichroism (XMCD) measurements on the disordered phase of Ba$_3$CoNb$_2$O$_9$ to determine the local electronic structure of the Co$^{2+}$ ions. We employed the line-fitting inverse partial fluorescence yield (iPFY) method to obtain reliable spectra from this insulating material. From the comparison between the experimental spectra and full multiplet cluster calculations we were able to constrain the local crystal-field environment and assess the spin--orbit-entangled nature of the Co$^{2+}$ ground state. In this chemically disordered system, the Co$^{2+}$ ions retain an effectively cubic crystal-field environment that supports a $J_{\mathrm{eff}} = 1/2$ ground state. These results further establish that Ba$_3$CoNb$_2$O$_9$ is a promising platform for investigating the effect of disorder on $J_{\mathrm{eff}} = 1/2$ magnetism.

\end{abstract}

\section{Introduction}

The emergence of unconventional magnetic states in correlated electron systems is often associated with geometric frustration, in which competing interactions suppress conventional long-range magnetic order. Triangular-lattice antiferromagnets have long served as prototypical examples, exhibiting strong quantum fluctuations and hosting a variety of nontrivial ground states
\cite{ANDERSON1973153,chubukovQuantumTheoryAntiferromagnet1991,balentsSpinLiquidsFrustrated2010,starykhUnusualOrderedPhases2015}.

Disorder and dilution provide another route to unconventional magnetic behavior. In site-diluted quantum magnets, randomness can fragment the magnetic lattice into orphan moments, finite clusters, and percolating spin networks, producing a broad distribution of local exchange and correlation energy scales. When combined with spin-$1/2$ quantum fluctuations, such disorder can give rise to dynamically fluctuating, short-range-correlated states without conventional static order or spin-glass freezing
\cite{sanaPossibleRealizationRandomnessdriven2024,hossainEvidenceRandomSpinsinglet2024,BCNO}.

\begin{figure}[h!]\centering\includegraphics[width=0.85\linewidth]{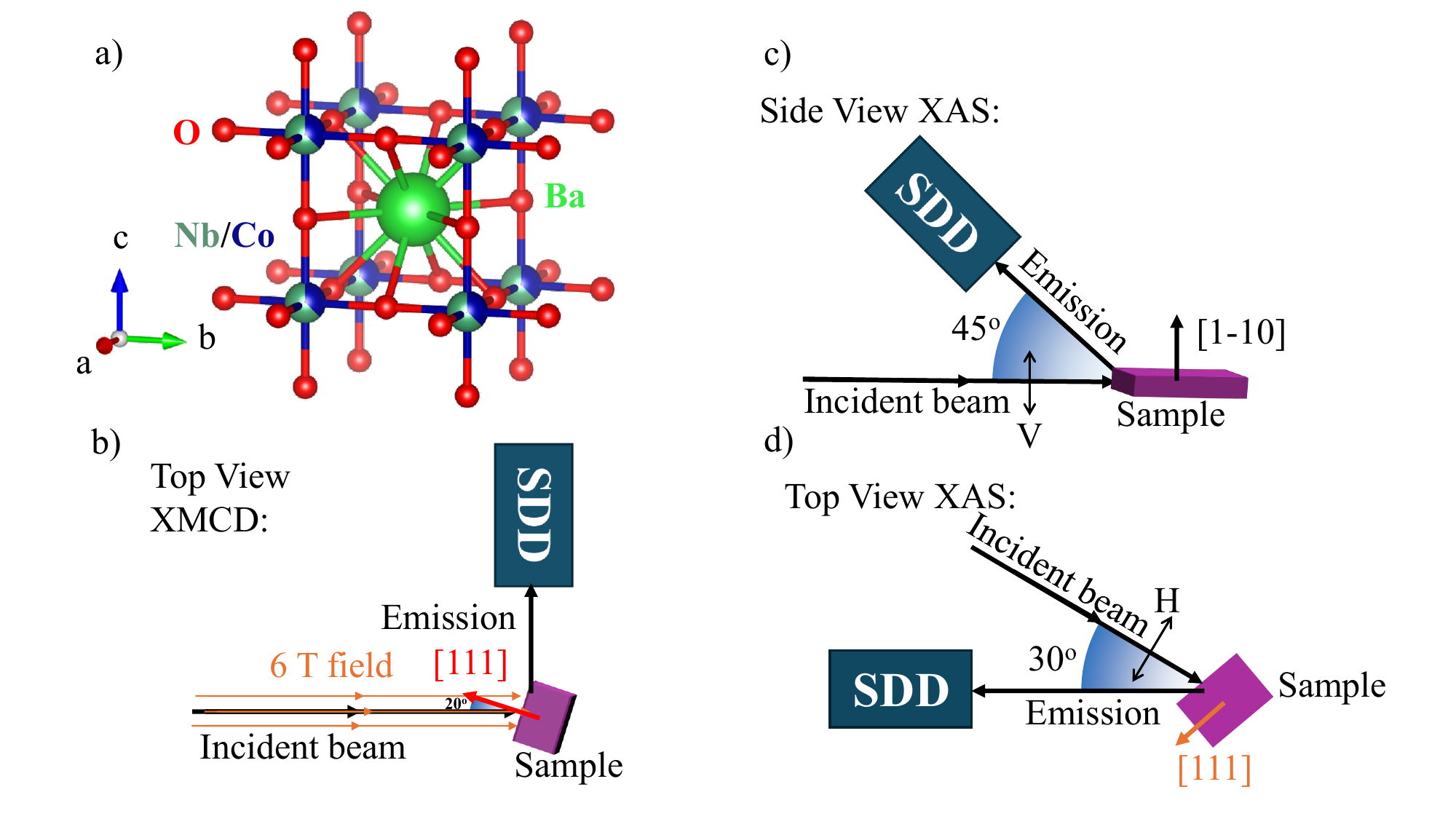}\caption{(a) Cubic crystal structure of Ba$_3$CoNb$_2$O$_9$ displaying inter-site mixing of Co and Nb. Schematic of the experimental geometry illustrating the orientation of the incident X-ray beam, polarization directions and the position of the silicon drift detector (SDD) for the (b) XMCD experiment, and for (c) and (d) the XAS experiment.}\label{fig:struc_geobcno}\end{figure}

Ba$_3$CoNb$_2$O$_9$ exhibits three polymorphs: K1, K2, and H, corresponding to the ordered cubic perovskite, the disordered cubic perovskite, and the hexagonal phase, respectively~\cite{treiberUberOrdnungsUnordnungsphanomeneBei1982}. The disordered K2 phase crystallizes in the cubic \textit{Pm$\bar{3}$m} space group, as shown in Fig.~\ref{fig:struc_geobcno}(a), and was first reported in the 1980s~\cite{treiberUberOrdnungsUnordnungsphanomeneBei1982}. Its formation depends sensitively on synthesis conditions: the K1-to-H transformation is irreversible, whereas high-temperature treatment of the H phase yields the disordered K2 phase~\cite{treiberUberOrdnungsUnordnungsphanomeneBei1982}.

Recent powder X-ray and neutron diffraction studies of the disordered phase revealed random occupation of the B sites by Co and Nb ions in a 1:2 ratio~\cite{BCNO}. Muon-spin rotation measurements found no evidence of static spin freezing down to 50~mK, while magnetic neutron diffraction refinements identified strong antiferromagnetic correlations between nearest-neighbor Co ions~\cite{BCNO}. The magnetic response was interpreted in terms of a diluted spin-$1/2$ Heisenberg model~\cite{BCNO}. These observations indicate that significant magnetic interactions persist despite site disorder.

A central question, however, is whether the assumed spin-$1/2$ degrees of freedom are microscopically justified in the disordered cubic phase. This point is particularly important because Co$^{2+}$ is not a simple spin-only ion. In octahedral coordination, Co$^{2+}$ has a spin--orbit-active $d^7$ electronic configuration, and its low-energy states are determined by the combined effects of the cubic crystal field, intra-atomic Coulomb interactions, covalency with the surrounding oxygen ligands, and spin--orbit coupling. These interactions give rise to a spin--orbit-entangled  Kramers doublet, but the precise character of this doublet can be sensitive to the local crystal-field environment.

Furthermore, Co$^{2+}$ compounds often possess low-lying excited states that can be thermally populated, making Curie--Weiss fits of the magnetic susceptibility unreliable for determining ground-state properties~\cite{burnusLocalElectronicStructure2008}. In the present compound, the random distribution of Co and Nb ions may also generate a distribution of local CoO$_6$ environments or weak deviations from ideal cubic symmetry. Such effects could, in principle, modify the crystal-field level scheme and the spin--orbit-entangled ground state. Establishing the local Co$^{2+}$ electronic structure is therefore essential for identifying the relevant low-energy magnetic degrees of freedom and for assessing whether the magnetism of disordered Ba$_3$CoNb$_2$O$_9$ can be described in terms of pseudospin-$1/2$ moments.

Here, we investigate the disordered phase of Ba$_3$CoNb$_2$O$_9$ using X-ray absorption spectroscopy (XAS) at the Co $L_{2,3}$ edges. XAS is an element-specific probe that is highly sensitive to the Co $3d$ electronic structure. Because dipole selection rules determine the accessibility and spectral weight of distinct final states, XAS, when combined with full-multiplet cluster calculations, provides direct information on both the ground state and low-lying excited states
\cite{burnusLocalElectronicStructure2008,tanakaResonant3d3p1994,degrootXrayAbsorptionDichroism1994,burnusValenceSpinOrbital2006,linLocalOrbitalOccupation2010,chinSpinorbitCouplingCrystalfield2019,ferreira-carvalhoTrigonalDistortionKitaev2025a}.

For Co$^{2+}$ in octahedral coordination, the temperature dependence of the Co $L_{2,3}$-edge XAS line shape is particularly sensitive to the thermal occupation of the low-lying $J_{\mathrm{eff}} = 1/2$, $3/2$, and $5/2$ states. By comparing temperature-dependent XAS spectra with full-multiplet calculations, we determine the energy splitting between these states and constrain the local crystal-field environment. We find that the Co$^{2+}$ ions in disordered Ba$_3$CoNb$_2$O$_9$ retain an effectively cubic crystal field, stabilizing a spin--orbit-entangled $J_{\mathrm{eff}} = 1/2$-like ground-state doublet. The resulting cluster model is further validated by comparison with XMCD spectra. These results provide a microscopic basis for treating the magnetism of disordered cubic Ba$_3$CoNb$_2$O$_9$ in terms of local Co$^{2+}$ pseudospin-$1/2$ degrees of freedom.

\section{Methods}

Single crystals were grown via the optical floating-zone method, with further details described elsewhere~\cite{BCNO}. XAS measurements were performed at the TPS~45A1 beamline of the National Synchrotron Radiation Research Center (NSRRC) in Taiwan~\cite{tsaiSubmicronSoftXray2019a}. The experimental geometry is illustrated in Fig.~\ref{fig:struc_geobcno}(c) and (d).

A single crystal was mounted on an aluminum sample holder and cleaved in a preparation chamber with a base pressure of $10^{-10}$~mbar to ensure a clean surface for the measurement. The sample was then transferred \textit{in vacuo} to the measurement chamber, which had a base pressure of $10^{-11}$~mbar. It was oriented such that the Poynting vector of the incident beam made an angle of 70$^\circ$ with the [111] direction.

XAS spectra were recorded at $T = 300$, 150, and 20~K. At 300~K, the incident soft X-rays were linearly polarized either horizontally (LH) or vertically (LV), whereas at lower temperatures only LH polarization was used. The LH polarization vector was 20$^\circ$ away from the [111] direction, while the LV polarization was parallel to the [$1\bar{1}0$] direction. A CoO reference sample was measured simultaneously upstream of the sample to allow precise incident-photon-energy calibration.

The iPFY technique~\cite{achkarBulkSensitiveXray2011,achkarDeterminationTotalXray2011a}, in its line-fitting variation~\cite{ferreira-carvalhoTrigonalDistortionKitaev2025a}, was used to obtain reliable XAS data. This method was employed since in Ba-containing oxides, the oxygen emission overlaps with the barium emission, preventing the Ba $M_{4,5}$ edges, which lie between the Co $L_{2,3}$ edges, from being measured properly. This overlap can lead to incorrect relative spectral intensities.

X-ray magnetic circular dichroism (XMCD) measurements were carried out at the ALBA Synchrotron~\cite{Bar2016}. The experimental geometry is shown in Fig.~\ref{fig:struc_geobcno}(b). The [111] direction was oriented 20$^\circ$ away from the Poynting vector of the incident beam. A magnetic field of 6~T was applied along the incident X-ray beam during the measurements. For each scan, the beam polarization was fixed to either right circular ($\sigma^+$) or left circular ($\sigma^-$) polarization to probe the circular magnetic dichroism. Spectra were collected at 150 and 30~K using the line-fitting iPFY method.

The XAS and XMCD spectra were calculated using full-multiplet configuration-interaction calculations~\cite{tanakaResonant3d3p1994,degrootXrayAbsorptionDichroism1994} for a CoO$_6$ cluster, as implemented in the Quanty code~\cite{haverkortMultipletLigandfieldTheory2012,haverkortBandsResonancesEdge2014,luEfficientRealfrequencySolver2014,Quanty}. The model includes intra-atomic Co $3d$--$3d$ and $2p$--$3d$ Coulomb interactions, atomic Co $2p$ and $3d$ spin--orbit couplings, O $2p$--Co $3d$ hybridization, and local crystal-field effects. The spin--orbit coupling constants and Coulomb multiplet parameters were obtained from Hartree--Fock calculations~\cite{cowanTheoryAtomicStructure1981}, while the monopole Coulomb terms and the O $2p$--Co $3d$ charge-transfer energy were estimated from photoemission studies of Co$^{2+}$ compounds~\cite{bocquetElectronicStructureEarly1996}.

Hybridization parameters for the full-multiplet configuration-interaction cluster were obtained from a tight-binding model based on an \textit{ab initio} Wannierized LDA band-structure calculation. The calculation was performed for a supercell containing three unit cells: one with Co and two with Nb. The DFT calculations were carried out using the FPLO code together with the FPLORE package~\cite{koepernikFullpotentialNonorthogonalLocalorbital1999,fplo21,fplore}. To improve agreement with the experimental spectra, the resulting hybridization parameters were scaled to 80\% of their original values.

To further characterize the Ba$_3$CoNb$_2$O$_9$ single crystals, we carried out X-ray diffraction measurements using Mo K$\alpha$ radiation on a Bruker D8 VENTURE diffractometer. The instrument is equipped with a bent graphite monochromator, providing an approximately threefold intensity enhancement, and a Bruker PHOTON CMOS large-area detector. The results are presented in the Appendix. 

\section{X-ray Absorption}

\begin{figure}[h!]
    \centering
    \includegraphics[width=0.8\linewidth]{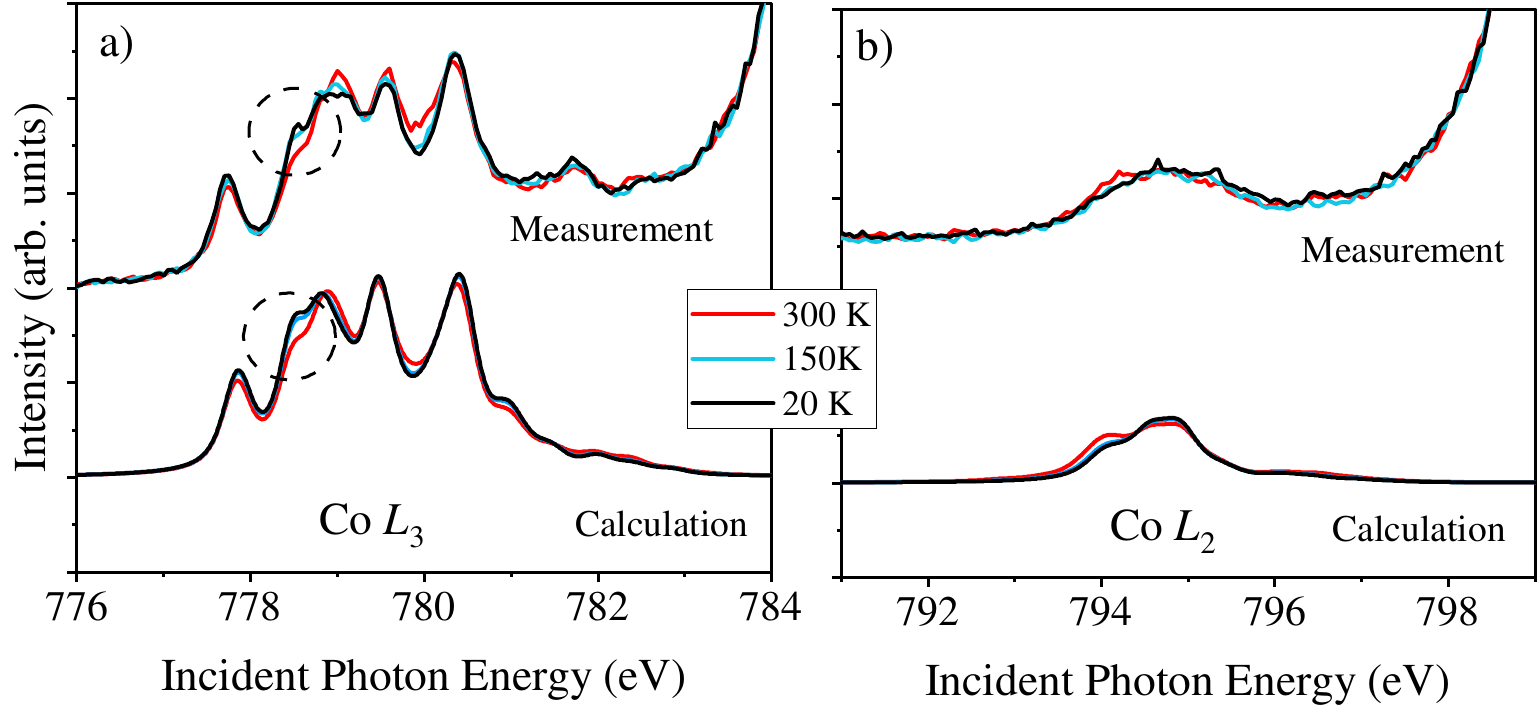}
    \caption{Comparison between experimental and calculated temperature-dependent Co $L_3$- and $L_2$-edge XAS spectra, shown in panels (a) and (b), respectively.}
    \label{fig:tldbcno}
\end{figure}

Figures~\ref{fig:tldbcno}(a) and \ref{fig:tldbcno}(b) show the temperature-dependent Co $L_3$- and $L_2$-edge XAS spectra of Ba$_3$CoNb$_2$O$_9$. The spectra exhibit the characteristic multiplet structure of high-spin octahedral Co$^{2+}$ \cite{degroot90,degrootXrayAbsorptionDichroism1994,burnusLocalElectronicStructure2008,ferreira-carvalhoTrigonalDistortionKitaev2025a}. Their temperature dependence reflects changes in the thermal population of low-lying spin--orbit-split excited states \cite{degrootXrayAbsorptionDichroism1994,ferreira-carvalhoTrigonalDistortionKitaev2025a}.

The Co $L_{2,3}$ spectral line shape is governed by a complex multiplet structure arising from intra-atomic Coulomb interactions, local crystal fields, and hybridization with oxygen ligands. To further understand the local electronic structure, we performed full-multiplet calculations for a CoO$_6$ cluster in cubic symmetry as a function of temperature. The parameters used in the calculations are listed in Ref.~\cite{parameters}. The good overall agreement between the calculated and experimental spectra supports the validity of the extracted electronic parameters.

Owing to the random distribution of Co and Nb ions, the measured spectra may represent an effective, spatially averaged response over different local Co environments. Such chemical disorder can produce a distribution of local crystal fields and may lower the local point symmetry experienced by individual Co ions~\cite{hauprichtLocalElectronicStructure2010}. 

\begin{figure}[h!]
    \centering
    \includegraphics[width=0.9\linewidth]{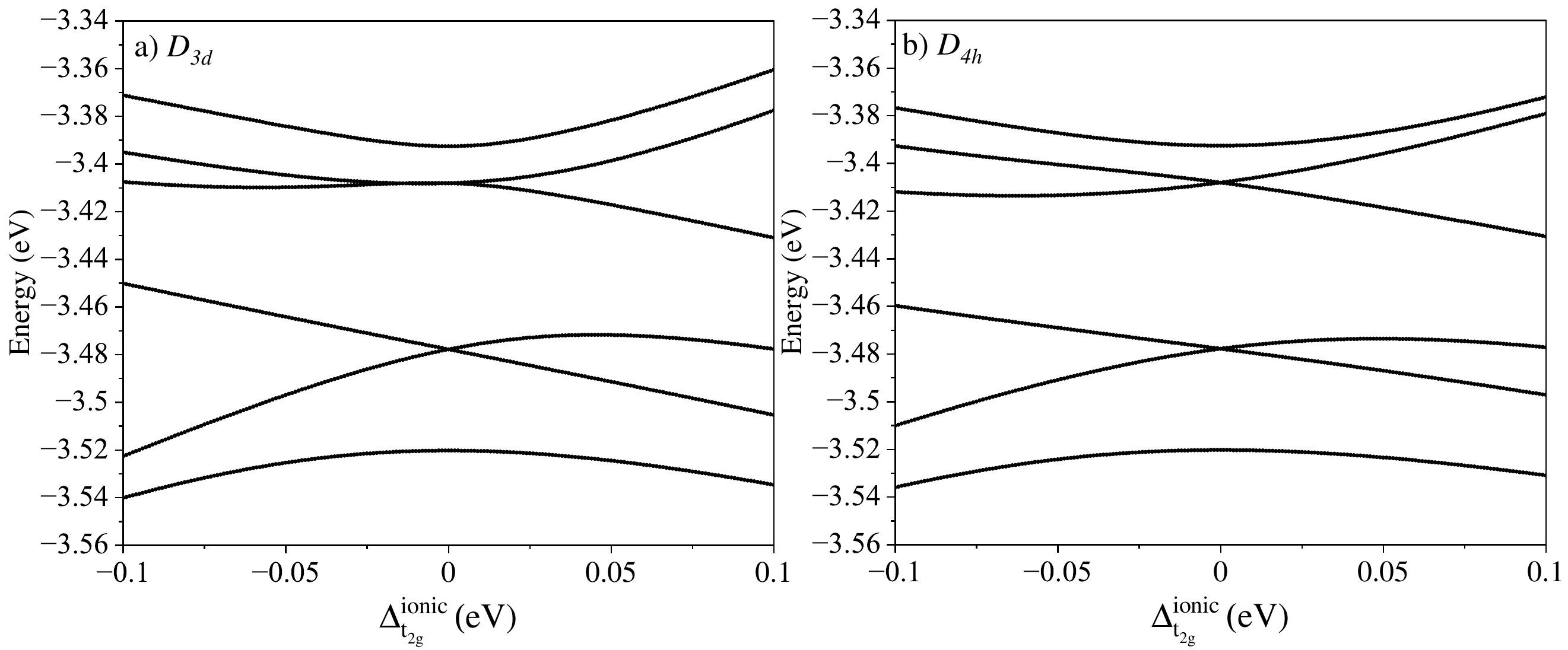}
    \caption{Energy level diagram of the Co $3d^7$ ion in the CoO$_6$ cluster showing the evolution of the lowest
    	12 eigenstates under the inﬂuence of trigonal ($D_{3d}$) and tetragonal ($D_{4h}$) crystal fields. $\Delta_{t_{2g}}^{ionic}$ indicates the ionic contribution to the crystal field splitting within the $t_{2g}$ subshell.}
    \label{fig:energylevel}
\end{figure}

To test for possible deviations from local cubic symmetry, we calculated the total energy level diagram of the CoO$_6$ cluster including trigonal ($D_{3d}$) and tetragonal ($D_{4h}$) distortions.  Fig.~\ref{fig:energylevel} shows that the energy separation between the ground state and the first excited state is about 42 meV for perfect octahedral coordination, i.e. in the absence of any trigonal or tetragonal distortions. This is reflected in the temperature dependence of the spectra as displayed in Fig.~\ref{fig:tldbcno} where the spectral feature at 778.5 eV energy, as highlighted by the dashed circle, shows first very little change when increasing the temperature from 20 K to 150 K, and then a significant change when increasing the temperature further to 300 K.

\begin{figure}
	\centering
	\includegraphics[width=0.8\linewidth]{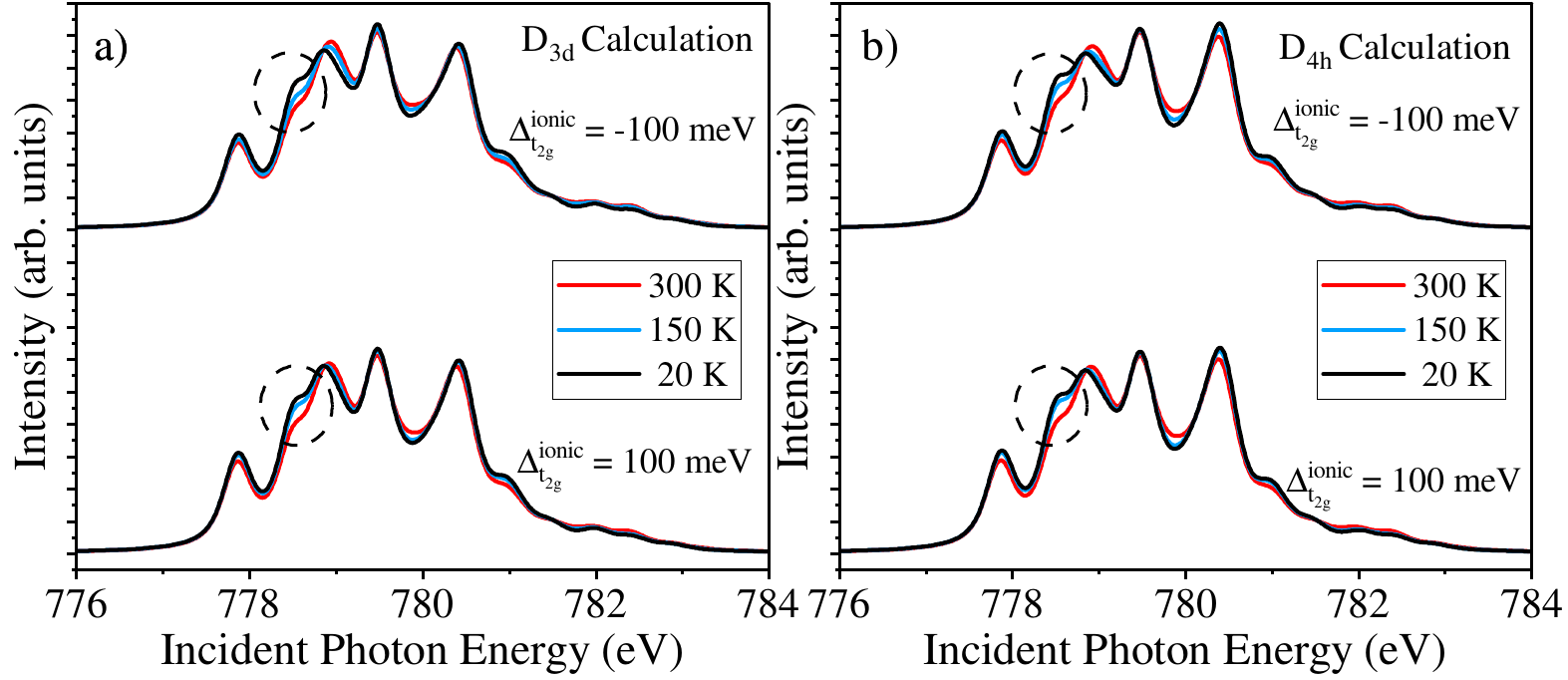}
	\caption{Calculated temperature-dependent Co $L_3$-edge XAS spectra of the CoO$_6$ cluster in the presence of trigonal ($D_{3d}$) and tetragonal ($D_{4h}$) distortions. The ionic contribution to the crystal field splitting within the $t_{2g}$ subshell is -100 meV and + 100 meV.}
	\label{fig:tdep_simulation}
\end{figure}

In the presence of a trigonal or tetragonal crystal field, the energy separation between the ground state and the first excited state decreases, as can be seen from Fig.~\ref{fig:energylevel}. For example, in $D_{3d}$ with -100 meV for the splitting within the $t_{2g}$ subshell, the first excited state is only 17.5 meV from the ground state. If such is the case, it will have observable consequences for the temperature dependence of the spectra. Indeed, the simulations in panel (a) of Fig.~\ref{fig:tdep_simulation} show that such a -100 meV crystal field will already show an appreciable change in the feature at 778.5 eV energy (highlighted by the dashed circle) when increasing the temperature from 20 K to 150 K, with a more modest change going to 300 K.  This is not what the experiment shows, see top curves of panel (a) in Fig.~\ref{fig:tldbcno}. The temperature effect on the spectra between 20 K and 150 K for a tetragonal crystal field or fields with opposite sign are more modest, see top curves of panel (b) and bottom panels of (a) and (b) in Fig.~\ref{fig:tdep_simulation}. Nevertheless, the experiment seems to show an even weaker temperature dependence. Perhaps one can conclude that if local crystal fields lower than octahedral are present in this disordered system, they are likely not larger than 100 meV. 

\section{X-ray magnetic circular dichroism}

\begin{figure}[h!]
    \centering
    \includegraphics[width=0.9\linewidth]{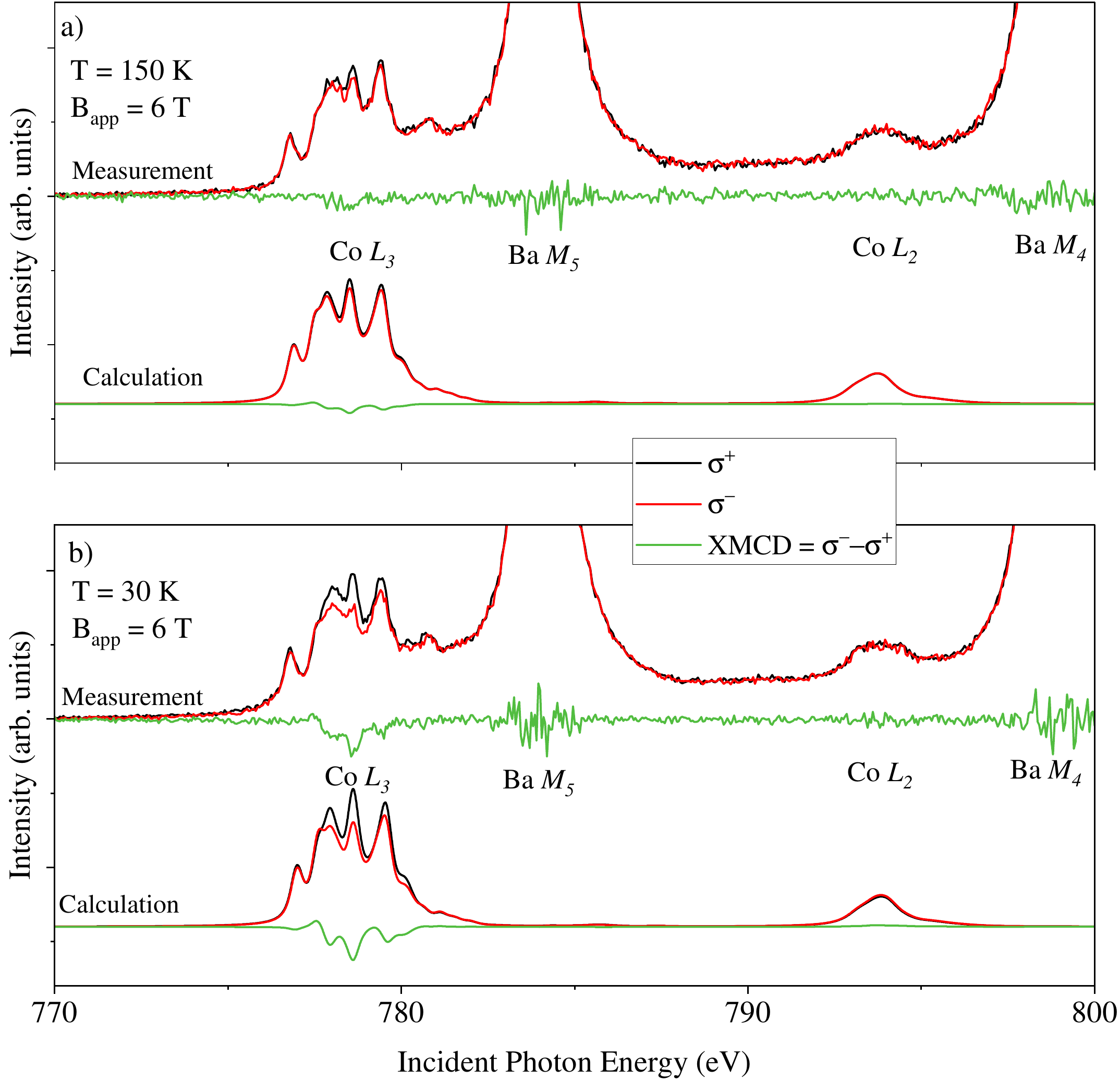}
    \caption{XMCD spectra of Ba$_3$CoNb$_2$O$_9$ \ measured under an applied magnetic field of 6~T. (a) shows the spectra recorded at 150~K, and (b) at 30~K. Both the experimental and calculated XMCD signals are defined as XMCD = left ($\sigma^-$) – right ($\sigma^+$). }
    \label{fig:xmcd_bcno}
\end{figure}

To further investigate the magnetic response of the Co$^{2+}$ ions, we performed XMCD measurements under an applied magnetic field of 6~T.

Figure~\ref{fig:xmcd_bcno} shows the measured and calculated XMCD spectra at the two measured temperatures. The spectra were recorded using circularly polarized light, $\sigma^-$ and $\sigma^+$, with a 6~T magnetic field applied along the Poynting vector of the incident X-ray beam. The XMCD signal was obtained as the difference between the spectra measured with opposite helicities, $\sigma^- - \sigma^+$.

The calculated spectra were obtained using the same parameters as those used for the XAS simulations within the cubic crystal-field model~\cite{parameters}, now including the external magnetic field of 6~T. It is important to note that magnetization measurements up to 25~T show that the magnetic moments are not fully polarized at the 6~T field used in our XMCD experiment~\cite{BCNO}. Consequently, the XMCD measurements probe a partially polarized magnetic state. To reproduce the experimental XMCD intensity, an effective exchange field of $H_{\mathrm{ex}} = -0.4$~meV, defined relative to the direction of the applied magnetic field, was included in the calculations. This term should be understood as an effective way to account for the incomplete polarization of the Co$^{2+}$ moments at 6~T, rather than as a microscopic determination of a specific exchange interaction.

From our XMCD calculation, we obtain for the ground state expectation values of $\langle S \rangle = -0.82$ and $\langle L \rangle = -0.63$, corresponding to an effective local magnetic moment of 2.26~$\mu_\mathrm{B}$/Co$^{2+}$. This value exceeds the magnetic moment of approximately 1.65~$\mu_\mathrm{B}$/Co$^{2+}$ measured at 25 T in bulk magnetization measurements~\cite{BCNO}.  Higher-field magnetization measurements would be valuable for clarifying the approach to saturation and the degree of moment polarization in this system. The good agreement between the calculated and measured XMCD line shapes therefore shows that the cubic cluster model captures the local electronic and magnetic structure of the Co$^{2+}$ ions, while the reduced bulk magnetization is consistent with collective effects beyond a single-ion description.

\section{Discussion}

Having established that the Co$^{2+}$ ions are best described by an effectively cubic crystal-field environment, we now analyze the corresponding ground-state wavefunction obtained from the cubic cluster model. The fit yields an effective cubic crystal-field splitting of $10Dq = 0.603$~eV. In the configuration-interaction calculation, the many-body ground state includes contributions from different charge-transfer configurations arising from hybridization between the Co $3d$ orbitals and the O $2p$ ligand states. The resulting ground-state wavefunction is written as:

\begin{equation}
\Psi(\mathrm{Co}^{2+}) = 
\alpha_7 |d^7\rangle 
+ \alpha_8 |d^8 \underline{L}\rangle 
+ \alpha_9 |d^9 \underline{L}^2\rangle,
\end{equation}
where $\underline{L}$ denotes a ligand hole and $\sum_{i=7}^{9}\alpha_i^2 = 1$. The fitted configuration weights are $\alpha_7^2 = 0.854$, $\alpha_8^2 = 0.143$, and $\alpha_9^2 = 0.004$, indicating a predominantly $d^7$ ground state with 2.86 holes in the Co $3d$ shell. These results further confirm that the ground state has a mainly Co$^{2+}$ character.

From the calculated ground-state wavefunction, we obtain an effective total angular momentum of $J_{\mathrm{eff}} = 0.833$. The deviation from the ideal $J_{\mathrm{eff}} = 1/2$ value reflects the fact that Co$^{2+}$ is a $d^7$ ion in a finite cubic crystal field, where the low-energy states cannot be described by a perfectly isolated $t_{2g}$ manifold. In the full multiplet description, intra-atomic Coulomb interactions, spin--orbit coupling, hybridization and the finite separation between the $t_{2g}$- and $e_g$-derived states produce residual admixture beyond the idealized $t_{2g}$-only limit. This admixture leads to a calculated $J_{\mathrm{eff}}$ value larger than the idealized $1/2$.

Nevertheless, this enhanced value does not imply a breakdown of the pseudospin description. The ground state remains a well-separated Kramers doublet derived from the spin--orbit-split ${}^4T_{1g}$ manifold, with the first excited state located at about 42~meV above the ground state. The ground state of this system can therefore still be regarded as an effective $J_{\mathrm{eff}} = 1/2$-like pseudospin doublet.

\section{Conclusion}

In this study, we establish a microscopic picture of the local electronic and magnetic properties of the disordered perovskite Ba$_3$CoNb$_2$O$_9$. Comparison of the experimental XAS and XMCD spectra with configuration-interaction cluster calculations revealed a predominantly $3d^7$ ground state with an effective octahedral crystal-field splitting of $10Dq = 0.603$~eV. The resulting ground state is a spin--orbit-entangled $J_{\mathrm{eff}} = 1/2$-like doublet, confirming that the low-energy magnetic degrees of freedom can be described in terms of Co$^{2+}$ pseudospin-$1/2$ moments. The presence of these entities makes Ba$_3$CoNb$_2$O$_9$ a compelling platform for investigating the effect of disorder on the pseudospin-$1/2$-based magnetism in $3d$ transition-metal systems.

%%%%%%%%%%%%%%%%%%%%%%%%%%%%%%%%%%%%%%%%%%%%%%%%%%%%%%%%%%%%%%
%%%%%%%%%%%%%%%%%%%%%%%%%%%%%%%%%%%%%%%%%%%%%%%%%%%%%%%%%%%%%%
%%%%%%%%%%%%%%%%%%%%%% ACKNOWLEDGMENTS %%%%%%%%%%%%%%%%%%%%%%%
%%%%%%%%%%%%%%%%%%%%%%%%%%%%%%%%%%%%%%%%%%%%%%%%%%%%%%%%%%%%%%
\section*{Data Availability}

The data that support the findings of this article are openly
available \cite{3.RDPRH9_2026}.

\appendix

\section{Appendix: Single Crystal X-ray Diffraction}

\begin{figure}[h!]
    \centering
    \includegraphics[width=0.7\linewidth]{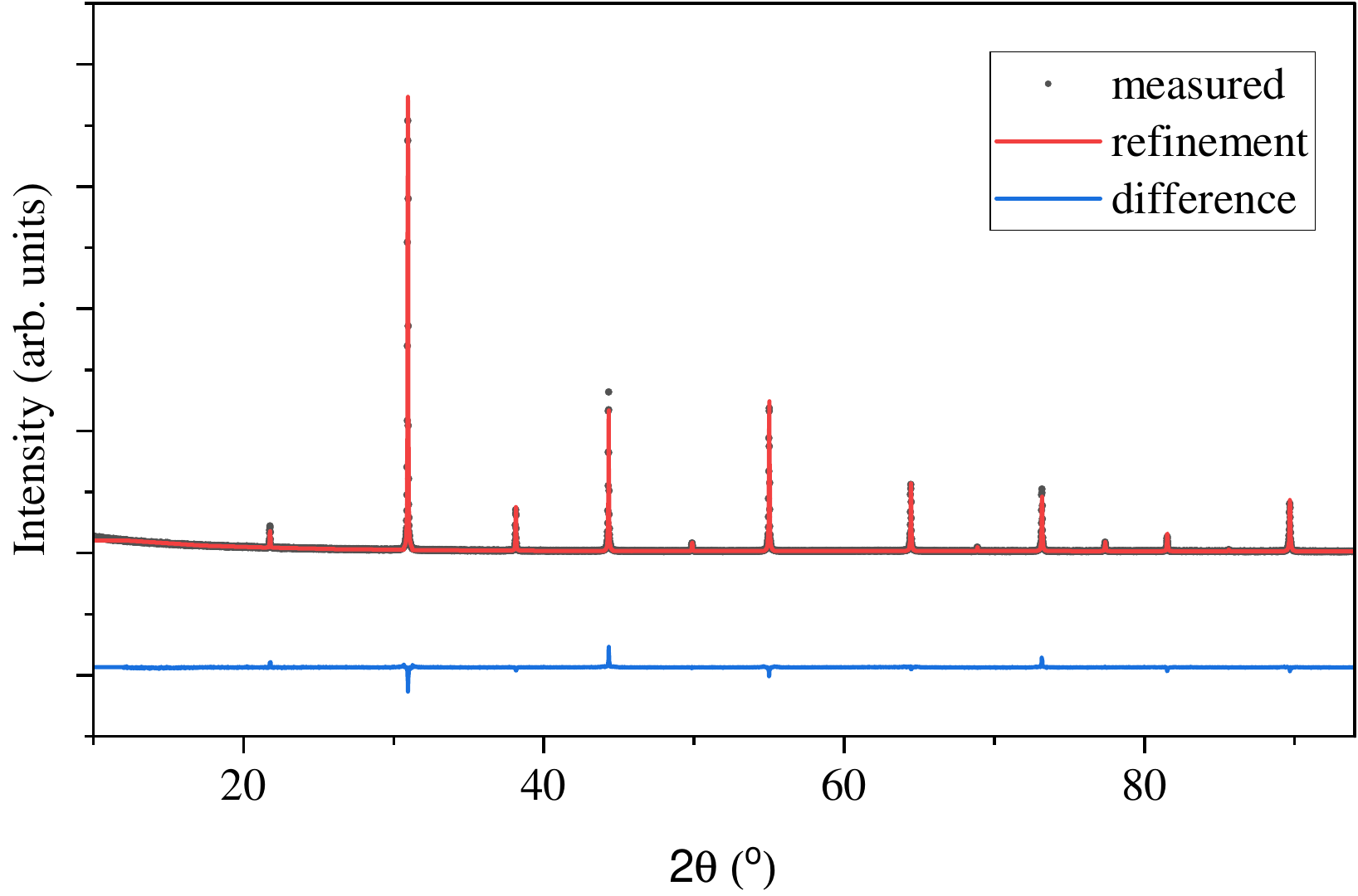}
    \caption{Room temperature powder X-ray diffraction pattern from a Ba$_3$CoNb$_2$O$_9$ single crystal. The black dots represent the observed intensities, the solid red line represents the calculated intensities obtained from a Rietveld fit. The solid blue line represents $I_\mathrm{obs} - I_\mathrm{cal}$.
}
    \label{fig:bcnoxrd}
\end{figure}

Using a  Mo K$_\alpha$ radiation, single crystal X-ray diffraction measurements have been performed on a Bruker D8 VENTURE diﬀractometer (with a bent graphite monochromator for about three times intensity enhancement and equipped with a Photon III CMOS large area detector) to study the crystal structure of our Ba$_3$CoNb$_2$O$_9$ sample. We studied three specimens, each approximately 20 $\mu$m in each dimension.
All measurements reveal the presence of a (disordered) cubic perovskite structure with space group \textit{Pm-3m} and with a statistical occupation of the B-site with Co and Nb.
The refinement results for one of the samples are listed in Tabs. \ref{tab:Tab1} and \ref{tab:Tab2} and the resulting
crystal structure is shown in Fig.~\ref{fig:struc_geobcno}(a).
The refinement of the B-site occupancies yields a Nb:Co ratio of 0.640(5) : 0.360(5) which is in good agreement with the expectation value of 2:1.
Complementary powder X-ray diffraction measurements with a high resolution setup (Bruker D8 Discover X-ray diffractometer with monochromatized Cu-K$_\alpha$ radiation) confirm these findings -- there are no further peak splittings nor additional reflections visible in these measurements which can be seen in Fig.~\ref{fig:bcnoxrd}.

\begin{table}
 % % \centering
  \begin{tabular}{l|l}
   \hline
Crystal   & Ba$_3$CoNb$_2$O$_9$ \\
density & 6.4781 g/cm$^3$ \\
Temperature & room temperature \\
Wavelength & Mo K$_{\alpha}$ \\
Crystal system & cubic \\
Space group & \textit{P m -3 m} (221) \\
Unit cell dimensions & $a$~=~4.0889(7) \AA \\ 
Volume               & 68.36(2) \AA$^3$ \\
Z & 1 \\
$F$(000)              & 116 \\
Crystal size  &   $\sim$20~$\mu$m  \\
2$\Theta_{max}$ &  144.38$^{\circ}$  \\
Index range   &  $h$: -9~$\rightarrow$~10 \\
              &  $k$: -10~$\rightarrow$~10 \\
              &  $l$: -10~$\rightarrow$~8  \\
Reflections in total / independent & 5249 /  175   \\
Observed reflections / independent & 5211 / 175 \\
R$_\sigma$ (obs/all) & 0.91\% / 0.91\% \\
R$_\mathrm{int}$ (obs/all) &   2.82\%\ / 2.82\% \\
Redundancy & 30 \\
Completeness up to $2$$\Theta$~=~144.38$^o$ & 98.31\%  \\
Absorption correction & multi-scan \\
Min. / max. transmission &  0.7780 / 1  \\
Refinement method & least squares on $F^2$ \\
Goodness of fit &  0.94 \\
R / R$_w$ ($I > 3\sigma(I)$) &  0.85\% / 2.29\%  \\
R / R$_w$ (all) &  0.85\% / 2.29\%  \\
\hline
 \end{tabular}
  \caption{Crystallographic \&\ structural refinement data of a single crystal X-ray diffraction measurement of Ba$_3$CoNb$_2$O$_9$.
    The crystallographic software \textit{Jana} was used for the structural refinement.}
\label{tab:Tab1}
\end{table}

\begin{table}
 % % \centering
  \begin{tabular}{l|cccccc}
   \hline
        \textbf{atom}  & \textbf{occ.}  & \textbf{x} & \textbf{y} & \textbf{z}    & U$_{iso}$ \\
   \hline
Ba1   &   1        &    0.5   &    0.5   &    0.5   &    0.00702(2) &  \\
Co1   &   0.360(5) &    0     &    0     &    0     &    0.00510(3) &  \\
Nb1   &   0.640(5) &    0     &    0     &    0     &    0.00510(3) &  \\ 
O1    &   1        &    0.5   &    0     &    0     &    0.00949(9) &  \\

   \hline \hline
        \textbf{atom}  & \textbf{$U_{11}$}  & \textbf{$U_{22}$} & \textbf{$U_{33}$} & \textbf{$U_{12}$}    & $U_{13}$ & $U_{23}$ \\
   \hline   
Ba1   &    0.00702(3) & 0.00702(3)  & 0.00702(3)  &    0   &    0   &    0 \\
Co1   &    0.00510(5) & 0.00510(5)  & 0.00510(5)  &    0   &    0   &    0 \\
Nb1   &    0.00510(5) & 0.00510(5)  & 0.00510(5)  &    0   &    0   &    0 \\
O1    &    0.0100(2)  & 0.00920(13) & 0.00920(13) &    0   &    0   &    0 \\

   \hline  
  \end{tabular}
  
  \caption{Refinement results of single crystal X-ray diffraction measurements of Ba$_3$CoNb$_2$O$_9$. $U_{i,j}$ are in units of \AA$^{2}$.\label{Tab2}}
\label{tab:Tab2}
\end{table}

\break
\ack{M.M.F.-C. thanks Oliver Stockert for valuable discussions.
M.M.F.-C. greatly acknowledges funding from the  Deutsche Forschungsgemeinschaft (DFG, German Research Foundation) Grant No. 387555779. Work at MPI-CPfS was partially supported by SFB1143 (Project No. 247310070). We would like to acknowledge the Core Lab Quantum Materials at Helmholtz-Zentrum Berlin for sample synthesis. The authors acknowledge the support from the Max Planck-POSTECH-Hsinchu Center for Complex Phase Materials. XMCD measurements were performed at ALBA under proposal 2023097766.}

\bibliographystyle{unsrt}
\bibliography{ref}

\end{document}